\documentclass[11pt,twocolumn]{article}
\usepackage{fancyhdr}
\usepackage{times}
\usepackage{amssymb}
\usepackage{amsmath}
\usepackage{bm}
\usepackage{array}
\usepackage[pdftex]{graphicx}

\newcommand {\blfootnote}[1]{{\xdef\@thefnmark{}~#1}}

\begin{document}
\thispagestyle{empty}

\twocolumn[
    \begin{@twocolumnfalse}
{\LARGE\bf  
Combinatorial coding in neural populations
}
\vskip 0.2in
{\large \bf
Leslie C. Osborne$^{a,*} $, Stephanie E. Palmer$^{b,*} $, Stephen G. Lisberger$^{a} $ and William Bialek$^{b} $
}
\vskip 0.1 in

$^{a} $Sloan-Swartz Center for Theoretical Neurobiology, Department of Physiology, and the Howard Hughes Medical Institute, University of California at San Francisco, San Francisco, California 94143.\\
\vskip -0.1in
 $^{b} $Joseph Henry Laboratories of Physics, and the Lewis-Sigler Institute for Integrative Genomics, Princeton University, Princeton, New Jersey 08544.\\
 \vskip -0.1in
 $^{*} $These authors contributed equally to this work.
\vskip 0.1in


To evaluate the nature of the neural code in the cerebral cortex, we have used a combination of theory and experiment to assess how information is represented in a realistic cortical population response. We have shown how a sensory stimulus could be estimated on a biologically-realistic time scale, given brief individual responses from a population of neurons with similar response properties. For neurons in extrastriate motion area MT, a combinatorial code, one that keeps track of the cell identity of action potentials and silences in individual neurons across the population, carries twice as much information about visual motion as does spike count averaged over the same group of cells.  The combinatorial code is more informative because of the diverse firing rate dynamics of MT neurons in response to constant motion stimuli, and is robust to neuron-neuron correlations. We provide a theoretical motivation for these observations that challenges commonly held ideas about the nature of cortical coding at the level of single neurons and neural populations. 
\vskip 0.1in


\end{@twocolumnfalse}
]

\setlength{\parindent} {1cm}

Our understanding of sensory representations in the cerebral cortex is built on two fundamental ideas, each of which emerged to some degree from the study of simpler systems. First is the concept of rate coding, that neurons respond to sensory inputs by changing the rate at which they generate action potentials or spikes$^\textrm{1-3}$. Second is the idea of feature selectivity, that neurons respond not to raw stimulus variables such as light intensity but rather to specific features such as spatial gradients and their orientation or motion$^\textrm{4-7}$. Many neurons in a small neighborhood of the cortex seem to have very similar feature selectivity$^\textrm{8,9}$, suggesting that averaging over this apparently redundant population is an essential component of the cortical code.

There is no question that neurons respond to sensory inputs by changing the rate at which they generate action potentials. However, the fact that neurons modulate their spike rate in response to sensory stimuli is a statement about their \textit{average} behavior in experiments where the same stimulus is repeated many times and responses are averaged to create the peri-stimulus time histogram, or PSTH. The brain itself, though, has no way of computing the average rate of an individual neuron, and all decisions must be based on single examples of the spike trains, albeit from a populaton of cells. Furthermore, many behaviors are guided by sensory information available in small time windows, so that each neuron can contribute only a handful of action potentials$^\textrm{10,11}$. This raises the critical issue of how the nervous system extracts information from such a small number of events.  

While ``rate coding'' is viewed as well established, codes based on the timing of spikes, whether in sequence from a single neuron or across a population, have been viewed as more speculative, except in special cases.  In particular, the fact that neurons in cortex generate spike trains that are approximately described by a modulated Poisson process means that the (time varying) rate provides a nearly complete description of the distribution out of which spikes are drawn, and this has been taken as \textit{prima facie} evidence against a timing code. We will argue that this informal inference from the statistics of spike trains to the structure of the neural code is incorrect, and we will reformulate the coding problem.  Rather than seeing the issue as ``rate codes'' vs. ``timing codes,'' we suggest that one can ask directly about the nature of the symbols that carry sensory information. 

Our paper begins by showing analytically that once rates vary as a function of time, the best estimate of rate from a single example of the spike train depends on the detailed timing of spikes. This effect is clear in experiments conducted in visual area MT as well, highlighting the limitations of the usual ``rate vs. timing'' formulation of the coding problem.  We then demonstrate the analogous effect in populations of neurons, showing that diversity in the dynamics of responses---even among neurons with nominally identical feature seelctivity---opens the possibility of a combinatorial code$^\textrm{10,12,13}$ in which stimulus features are represented in patterns of spiking and silence across a population of MT neurons. Even in populations of modest size ($N\sim 20$ cells), these patterns provide more than twice the amount of information about the stimulus than is available from pooling the spike counts. While we do not know if the cortex actually uses additional information in patterns of spikes and silence across the population, we show that the additional information does not require any unusual properties of the neural spiking statistics, and thus could exist in almost any population of cortical neurons.

\noindent \textbf{\textit{Results}}

\noindent \textit{{Relationship of firing rate and spike timing for Poisson neurons}}

The time course of the firing rate of a neuron can be estimated by accumulating a peri-stimulus time histogram (PSTH) across multiple responses to the same stimulus or motor response.  In reality, however, the nervous system does not have the opportunity to estimate the underlying rate of a neuron's response by averaging across multiple nearly-identical behavioral epochs.  If it does estimate the time-modulated rate of a neuron, $r(t)$, then it must do so on the basis of one sequence of all or nothing events at specified times. 

To see how this would work, consider a neural spike train that is a modulated Poisson process and assume that we observe the spike train in a window of time $0<t<T$. For a Poisson neuron with rate $r(t)$, the probability density for spikes to occur at times $t_1, t_2, \ldots, t_n$ in the window is given by  
\begin{multline} \label{GrindEQ__1_} 
 P\left [ t_1,t_2,\ldots t_n\right |r(t)] = \\ \frac{1}{n!}\exp\left [ -\int_0^Tdt r(t)\right ] r(t_1)r(t_2)\ldots r(t_n).
\end{multline} 
To estimate the rate $r(t)$ from observations on the spike train, we use Bayes' rule to construct the probability distribution of rates given our observations: 
\begin{equation} \label{GrindEQ__2_} 
P\left [ r(t) | t_1,t_2,\ldots t_n\right ] =\frac{P\left [ t_1,t_2,\ldots t_n\right |r(t)] P\left [ r(t) \right ] }{P\left [ t_1,t_2,\ldots t_n\right ] }
\end{equation} 
where $P[r(t)]$ is the probability distribution for the rates $r(t)$ accessed by the neuron over its dynamic range of responses, and $P[t_1,t_2,\ldots t_n]$ is the total probability of observing this sequence of spikes, averaged over stimuli. If $\bar{r}=\frac{1}{T} \int _{0}^{T}dtr(t) $ is the average of this rate over the whole window $T$, then some alegbra reveals that: 
\begin{equation} \label{GrindEQ__3_} 
P(\bar{r}|t_1,t_2,\ldots t_n) \propto \exp (-T\bar{r})\langle r(t_1)r(t_2)\ldots r(t_n)\rangle_{\bar{r}}
\end{equation} 
where $\langle\ldots\rangle_{\bar{r}}$ denotes an average over all the functions $r(t)$ used by the neuron that have the same average value $\bar{r}$. The important aspect of Equation \ref{GrindEQ__3_} is that the timing of the individual spikes has not disappeared from the result; to estimate the underlying spike rate from a single response, we need to know the timing of the spikes. 

\begin{figure}
\begin{center}
\leavevmode
\hbox{
{\includegraphics*[width=2.3in]{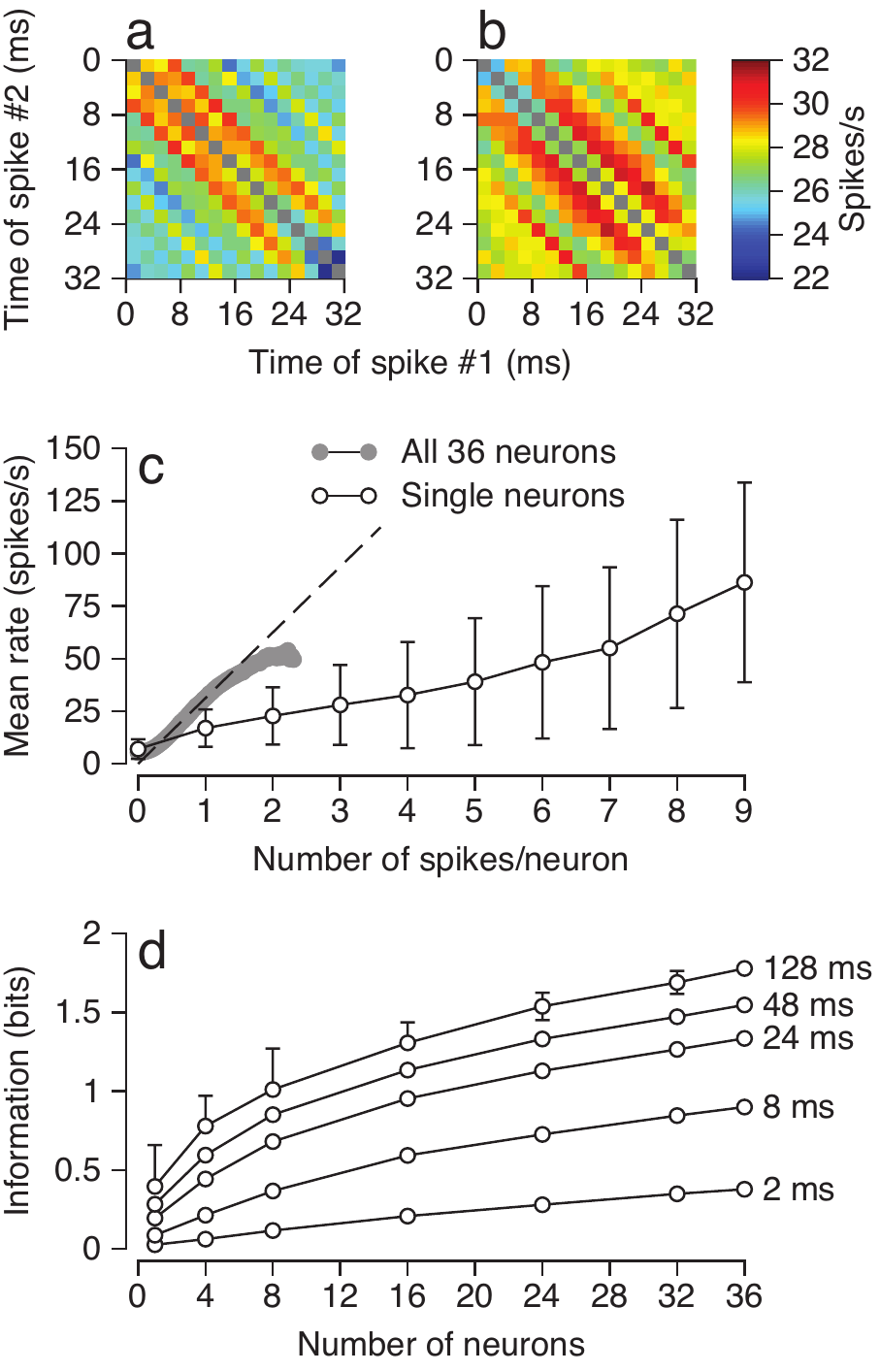}}
}
\caption{Relationship between underlying rate and spike count for single neurons and neural populations. (a) Relationship between spike times and the underlying mean rate for 32 ms analysis windows that contain two spikes, based on the data from a single MT neuron. The times of the two spikes are indicated on the abscissa and ordinate, and the color code indicates mean rate. (b) Same as (a), but after the spikes in each time window had been shuffled across trials. (c) Conditional mean rate, the average of observed rates given a particular spike count in a 32 ms analysis interval, plotted as a function of the number of spikes in the window.  Symbols show rate estimated from single neurons individually then averaged over our sample of 36; gray ribbon shows rate estimated by counting spikes across single trial draws from each of the 36 neurons (pooled count). Error bars on the one cell pool indicate standard deviations of the mean over the 36 cells recorded. Dashed line shows the na\"ive rate, the number of spikes indicated on the abscissa divided by the duration of the analysis interval, $n/T$. (d) Information about underlying rate from spike counts as a function of the number of cells in the population. Different curves show calulations based on different duration analysis intervals, indicated by numbers to the right of each curve.}
\end{center}
\end{figure}

Figure 1 amplifies the significance of Equation \ref{GrindEQ__3_} on the basis of recordings from our sample of 36 MT neurons.  For each neuron, we measured the underlying rate by averaging across many trials to create PSTHs with a  time resolution of 2 ms. We then analyzed each trial individually, taking all 32 ms windows in which there were $n=1,2,\ldots,9$ spikes and averaging the underlying rates associated with each spike count.  In Figure 1a, we consider windows that contain $n=2$ spikes, and ask whether the rate in this window, defined as an average over trials, was related consistently to the timing of the spikes.  The variation in the color across the two dimensional map indicates that spike timing was related to underlying rate in a complex way.  To ask whether the same complex relationship would appear when the within-trial structure of spike timing was abolished, we randomized the trial identity of the spike train independently within each 2 ms time window. The modulation of rate across the map in Figure 1b shows that the relationship between spike timing and underlying rate persists, as predicted by Equation \ref{GrindEQ__3_}, even when we enforce the Poisson nature of the spike train and eliminate any correlations across time.  Similar patterns appear when the same analysis is performed on data from Poisson model neurons with sinusoidal modulations of the underlying firing rate (data not shown).  

We emphasize that these results about the role of timing in the estimation of rate are contrary to a widely held intuition, namely that for Poisson processes counting spikes in a window provides the best estimate of the underlying rate.  This is exactly correct for constant rates, but once rates vary in time the estimation problem changes its structure. To explore this further, we look in more detail at the relationship between the spike count in a single trial and the underlying rate as estimated by the PSTH.  Although it is well known that, especially when the counts are small, there will be significant random errors in estimating the rate, Figure 1c (open circles) shows that there are also large systematic errors. For almost all counts, the average rate in windows with a particular spike count $n$ falls far below the na\"ively expected value of spike count divided by analysis window duration $(n/T)$. When we pool spikes from across our full sample of 36 neurons (Fig. 1c, filled circles), the rates are much closer to the ``count per time'' estimate, but only over a highly restricted dynamic range.  

 To assess the cost of this reduced dynamic range, we ask directly how much information the spike counts provide about the underlying rates in our sample of data from MT.  Increasing either the time window for counting or the number of neurons used in the analysis increased the information about rate from counts, as shown in Figure 1d.  But these potential gains are constrained by the time scales of behavior: for the example of smooth pursuit eye movements, which are driven by the population responses in MT$^\textrm{14}$, the time window of analysis of visual motion is approximately 25 ms$^\textrm{15}$. In a 24-ms analysis interval (Fig. 1d), even pooling across 36 neurons allowed spike counts to provide much less than 2 bits of information about the underlying mean rate of the population, meaning that only 4 different values of rate can be distinguished perfectly. In contrast, the trajectory of the smooth pursuit behavior itself provides roughly 12 bits of information about the parameters of target motion$^\textrm{16}$.  While the brain probably pools over more than 36 neurons, these results certainly suggest that we should explore other possible coding schemes.

\begin{figure*}
\begin{center}
\leavevmode
\hbox{
{\includegraphics*[width=5in]{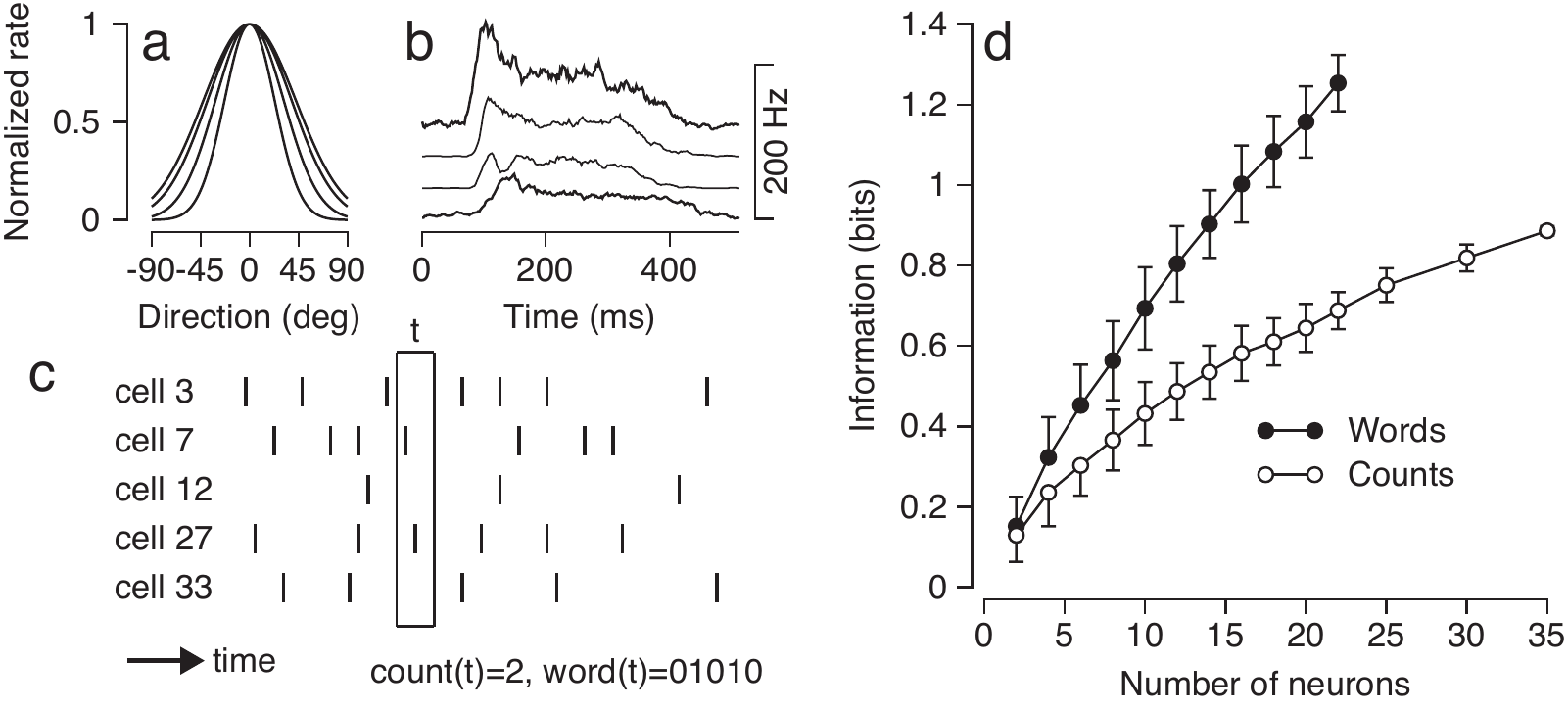}}
}
\caption{Utility of patterns of spiking and silence across a diverse population of MT neurons for providing information about a target motion stimulus. (a) The normalized tuning curves of four MT neurons, showing firing rate versus direction of motion. Data were normalized by the response to the preferred direction, relative to which all other directions are measured. (b) Responses of the same four neurons, plotting rate (from PSTH) versus time in response to a 256 ms step of target speed in the preferred direction. Firing rate curves have been offset horizontally to improve visibility. (c) Method for creating words to indicate spiking and silences across a population of neurons, in 8 ms windows. In this example, the population response at time ``t'' is characterized by the word: ``01010''. (d) The information that counts and words carry about the visual motion stimulus plotted as a function of the number of neurons in the analysis population.}
\end{center}
\end{figure*}

\noindent \textit{{Extra information about stimulus properties from patterns of spikes}}

The coexistence of time-dependent Poisson firing with the importance of spike timing in single neurons has an analog in populations of neurons.  Thus, if we have a group of cells in which rates vary across the population, then the combinatorial patterns of spikes and silence in a given small window of time may provide extra information beyond that available from pooling and counting the total number of spikes.  What is critical for thinking about this possibility in the cortex is that even neurons with similar feature selectivity can have very different responses dynamics, enabling combinatorial coding even in a population of nominally redundant cells.

To evaluate the possible utility of a combinatorial code, we consider a population of  MT neurons.  Experimentally, we have observed many responses to each of a finite set of different stimuli$^\textrm{11}$. Each of the cells in our sample was directionally tuned, with relatively similar selectivity and bandwidth when responses are normalized (Figure 2a), although the population had a wide range of maximal responses. Further, in response to a step of stimulus motion at the preferred speed and direction, different neurons showed considerable diversity in the dynamics of their firing rates (Figure 2b). If we assume that each neuron responds independently to its sensory inputs, then we can draw a single trial response from each neuron in our data base to create a model population response, even though the samples were recorded sequentially from many different neurons. Pooling the draws in different ways creates many different hypothetical neural populations of different sizes (see Methods for details). We can then subject each draw to various analyses to evaluate the possible nature of the neural code in a population response.  Finally we create correlated populations and evalute their impact on different coding schemes.  

If we look in a small window of time $\Delta\tau$, then the $i^\textrm{th}$ cell generates $n_\textrm{i}$ spikes, with $i=1,2,\ldots , N$. For small values of $\Delta\tau$, we will almost never see two spikes from a single cell.  Thus, the response of the population $\{ n_\textrm{i} \} $ can be treated as an $N$-letter binary word, $w$ (a pattern of 1's and 0's), as shown in Figure 2c. By keeping track of the combinations of spiking and silence across the population, we can ask how much information these code words carry about the stimulus. At each instant of time, the stimulus in our experiments is specified by the direction of motion $\theta $ and the time, $t-t_\textrm{onset} $, since the onset of motion, and calculations described in the Methods allow us to use the experimental data to estimate the information that the number or pattern of spikes provides about the stimulus, $I(w;\theta,t=t_\textrm{onset})$. 

The results in Figure 2d demonstrate that the information provided by binary code words increases as a function of the number of cells  that contribute to the word, exceeding one bit for a population of 16 neurons. If we use the same draws from the experimental data to estimate the information that the spike count $n\equiv \sum _{{\rm i}=1}^{N} n_{{\rm i}} $ provides about the stimulus, $I(n;\theta ,t-t_{{\rm o}nset} )$, we find that the total amount of information from spike counts is smaller than the total information from words, and never exceeds 1 bit even when all  neurons in the sample are pooled to obtain spike counts.  The combinations of of spiking and silence in this model population provide more than twice as much information as the pooled spike counts, even though the cells we are pooling from have nominally redundant feature selectivity.

To ascertain which feature of the neural response was responsible for the extra stimulus information available from words versus pooled spike counts, we next created a number of carefully contrived populations of 10 model Poisson units that preserved either the diversity of time varying firing rates or the diversity of direction tuning curves, or that eliminated all diversity (see Methods).  For each population, we then performed the same set of information calculations that led to Figure 2d.  For a population of model units that preserved the diversity of firing rate dynamics, $r(t)$, but forced all the neurons to have the same direction tuning curve (Fig. 3a, filled circles), the amount of extra information from words was the same as that for the draws from the experimentally observed spike trains of MT neurons (open circles).  If we contrived each unit to have the same time-varying trajectory of firing rate $r(t)$, but retained the diversity of directional tuning amplitudes, then about half of the extra information from words was lost (open triangles).  The extra information that remains reflects the fact that tuning curve diversity imposes different time-averaged absolute firing rates across the population, even if the trajectory of the trial-averaged firing rate $r(t)$ was the same for each model neuron.  We note that similar results on this latter point were obtained by Shamir and Sompolinsky, examining the effects of simulated heterogeneities in static tuning on population codes$^\textrm{17}$. Finally, if we created populations of fully redundant model units with one uniform trajectory, $r(t)$, and the same amplitude and width of direction tuning curve, then the extra information from words was lost, as expected (filled triangles). Analysis of information as a function of time revealed that the extra information from words was concentrated near the time of the onset transients of the neural response, where the diversity of response dynamics is greatest (data not shown).

\begin{figure}
\begin{center}
\leavevmode
\hbox{
{\includegraphics*[width=3.2in]{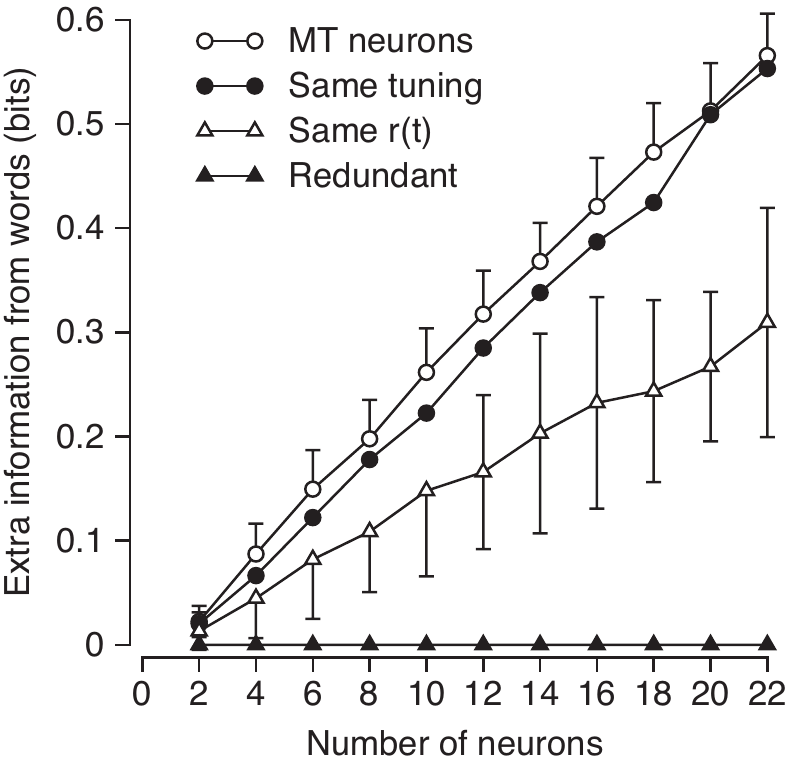}}
}
\caption{Combinatorial coding is enabled by a diversity of response dynamics. Extra information from words versus counts is plotted as a function of the number of neurons in the analysis population. Different symbols show the results from populations of real and model neurons with different features made redundant.  Open circles, data drawn from actual single trial responses; filled circles, diversity of response dynamics in model neurons mimics that in the data but each neuron has been made to have the same time-averaged direction-dependent response amplitude (i.e. same tuning curve); open triangles, model neurons that have the same time-varying firing rate, but response amplitude varies as in the actual data; filled triangles, model neurons that have the same time varying firing rate and direction-dependent response amplitudes but are independent Poisson processes.}
\end{center}
\end{figure}

\noindent \textit{{What does the extra information tell us?}}

The results of of the previous section tell us\textit{ how much} information the patterns of spiking and silence can convey about the stimulus. The next step is to understand \textit{what} these patterns are telling us about the stimulus. To focus our attention on a manageable set of patterns, we first computed the information carried by words and counts for different total spike counts in populations of $N=2\ldots 16$ cells, drawn 100 times from our 36 cortical neurons. Figure 4a shows that most of the extra information carried by words vs. counts comes from those words with relatively few spikes, that is from analysis windows when only a few of the neurons in the population emitted spikes and the rest were silent. Further, most words had zero, one, or two spikes and increasing numbers of spikes were progressively less common (Figure 4b). Combining these two effects shows that the dominant term in the extra information provided by words typically comes from instances when only one neuron fired a spike. Even when the size of the population was increased to 16, most of the extra information still arose from words of only one or a few spikes (Fig. 4c).  To understand what features of the stimulus are represented by different binary words, we therefore focused on words with only one spike. 

\begin{figure*}
\begin{center}
\leavevmode
\hbox{
{\includegraphics*[width=4.2in]{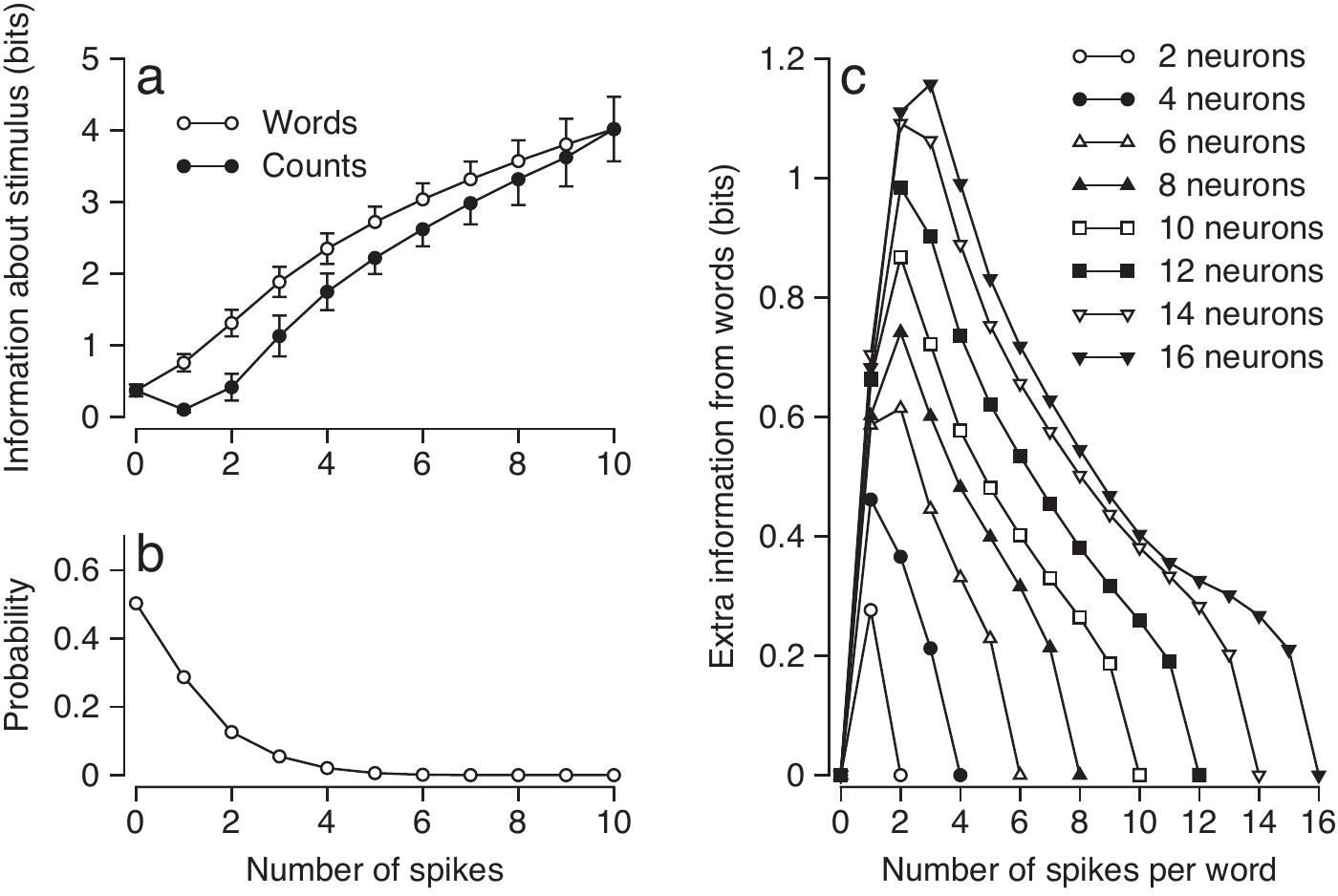}}
}
\caption{Added information from words analyzed separately for each spike count.  (a) Comparison of information from patterns of spiking and silences versus counts as a function of the number of spikes in the analysis window. Information from words was calculated by averaging the information from words of a given count. (b) The probability of observing each given spike count in 100 populations of 10 MT neurons. (c) Information from words minus information from counts is plotted separately for each spike count and each number of neurons in the analysis population.  Connected sets of symbols show data for all counts in a given population size.}
\end{center}
\end{figure*}

Our next step was to construct the response conditional ensembles$^\textrm{18}$, the distribution of stimuli that were associated with a particular neural response. We can think of these ensembles as ``receptive fields'' for the population response defined by the occurrence of a particular pattern of spiking and silences, and of the process used to create them as a population word variant of spike-triggered averaging. In the color maps of Figure 5, the color of each pixel shows the probability of a given direction of stimulus motion at a given time between motion onset and the time of the word in question. The responses were assembled across all stimulus motions and all times for a sample of 9 neurons. The occurrence of $n=1$ spike in a population of $N=9$ neurons is highly ambiguous in terms of the stimulus that elicited it, as seen in Figure 5a, where the red ring shows the wide range of stimulus directions and times that had high probabilities for a count of 1 spike.

\begin{figure}
\begin{center}
\leavevmode
\hbox{
{\includegraphics*[width=2.9in]{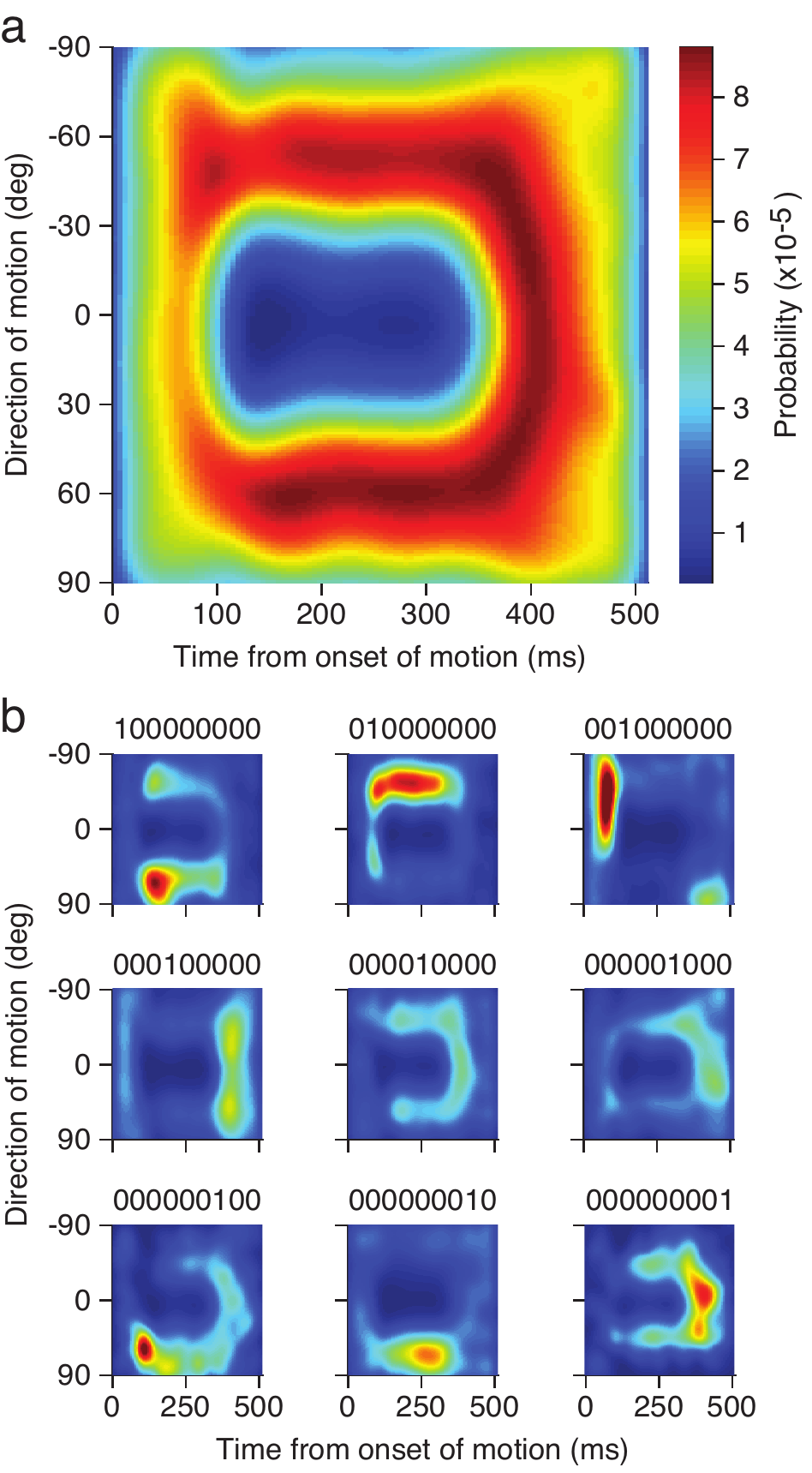}}
}
\caption{Response conditional stimulus ensembles for binary code words corresponding to a spike count of $n=1$ in a population of $N=9$ neurons. (a) The distribution of directions of motion and delays from motion onset, $P(\theta ,t-t_\textrm{onset} | n)$, given that the population of cells produced a total of one spike in window of size $\Delta \tau =8$ ms. (b) The same analysis, but now performed separately for each combination of spiking and silence where one neuron emitted a spike and all the others were silent. The probabilities in (a) have been normalized so that the total probability in the square is unity, with red representing the highest and dark blue the lowest values. The distributions in the small panels are normalized so that the average of all nine small panels yields the distribution in (a).  Graphs are based on analysis of draws from actual data in one group of 9 MT neurons. $N=9$ was chosen to allow the 3x3 presentation.}
\end{center}
\end{figure}

The event that contains one spike from nine neurons is composed of nine possible binary words, from 10000000 through 000000001, in which each single neuron spikes and all others are silent. Figure 5b shows that each binary word points to a different distribution of stimuli, and that each word actually represents a quite narrow range of stimulus directions and times from motion onset. Importantly, the binary words go a long way toward resolving the ambiguity between motion direction and motion onset time that is present in the neural data analyzed for Figure 5a, but not in a behavior driven by the MT population response, namely smooth pursuit eye movements$^\textrm{19}$.  Notice that if the neurons really were redundant, as one might have thought from their tuning curves, each of the events would have to point to the same distribution of stimuli, and each word would be associated with a distribution identical to that found by counting the total number of spikes.  The extra information in 1-spike words versus counting 1 spike is a general property of our cortical population and a similar plot to Figure 5 could be constructed for any group of cells.

Figure 5 demonstrates that different patterns of activity across a population of MT neurons can represent different stimuli, but not necessarily that the\textit{ combination} of spiking and silence is telling us anything that the spikes alone do not. We tested this directly (Fig. 6) by constructing a set of response conditional ensembles based on keeping track of the spike from a single cell and progressively discarding knowledge of silence in other cells. While the combination of spiking in neuron \#1 and silence in the rest of the population (100000000, upper left color map) points to a specific, small area in the space of stimuli, specificity declines in the representation as we throw away the knowledge of silence in more and more cells. Finally, the occurrence of a spike in neuron \#1 with no knowledge about the state of the other cells (1********, lower right color map) points to a large area with tens of degrees of uncertainty about motion direction and hundreds of milliseconds of uncertainty about the time of motion onset. In the example illustrated in Figure 6, it is striking that the most uncertain large blob in the lower-right panel of Figure 6 has almost no overlap with the original distribution of stimuli conditional on spiking in neuron \#1 and silence in the others: combinations of spikes and silence not only carry more information than spike counts alone, but they also stand for very different events in the sensory input. Synergy of spikes and silence was a common feature of our MT data, as observed previously in the retina$^\textrm{20}$.  For 10-cell groups, approximately 30\% of all 1-spike words have significant spike-silence synergy.  The prevalence of synergy increases with $N$: more than 60\% of 16-cell, 1-spike words are synergistic.  

\begin{figure}
\begin{center}
\leavevmode
\hbox{
{\includegraphics*[width=3in]{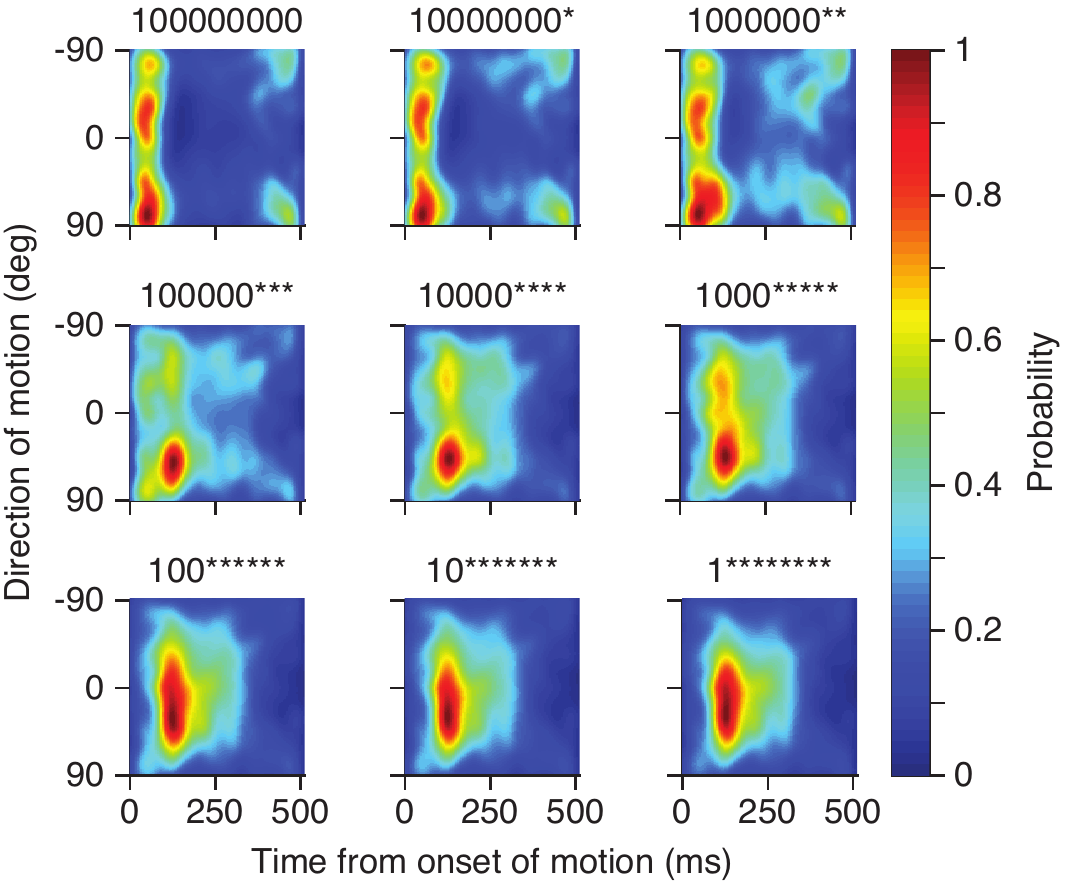}}
}
\caption{Contributions of silences in other neurons to the distribution of stimuli conditional on a spike in one neuron. Each pixel indicates the probability of a given direction of target motion at a given time after the onset of target motion, given a particular word of spiking and silences across the population of 9 MT neurons. The string above each color map indicates the word that was used to create each response conditional ensemble, where a ``1'' or ``0'' indicates the presence or absence of a spike in a neuron and a ``*'' indicates a wildcard so that an interval was included in the average whether a spike was present or absent. Further analysis revealed that the  100000000 and 1******** words contains 0.71 and 0.44 bits of information, respectively, about the stimulus.}
\end{center}
\end{figure}

\noindent \textit{{Effect of neuron-neuron correlations}}

So far, our discussion of the population responses in MT has assumed that the cells respond independently to sensory inputs. We ignored correlations between the responses of different neurons not just for simplicity, but also to give the classical model of averaging over multiple redundant cells the greatest chance to succeed. We found that the diversity of temporal dynamics in neuron responses makes a substantial change in the structure of the problem, opening the possibility for a form of combinatorial coding. We now ask whether correlations among neural responses alter the utility of a combinatorial code. 

Suppose that we know the average correlation coefficient between pairs of cells in a neural population.  We would like to construct model population responses that are consistent with this level of correlation, and of course also with the observed time dependent firing rates.  There are many ways to construct correlated populations, some of which correspond to complicated patterns of correlation which will give an obvious advantage to combinatorial codes.  To avoid this, we used the one parameter model described in the Methods to achieve a model population with a predefined mean level of correlations, but with a distribution that is otherwise is as random as possible.  The parameter in our model is a ``coupling,'' $J$ (see Equation 11) that we varied systematically to control the average pairwise correlation, which we assessed for each model population.  We then computed the statistics of the model population responses with different levels of mean correlation, and examined the information content of these responses, as before. 

In Figure 7, we illustrate the impact of correlations on the information encoded by populations of $N=10$ neurons. As expected from prior work$^\textrm{21-27}$, the information available from counting spikes is reduced when we add positive correlations among cells because it increases the trial-by-trial variance of the spike counts we obtain by summing across the neurons in a population. In contrast, negative correlations reduce the count variance and enhance information transmission. For coding based on patterns of spiking and silence, small positive correlations also cause a slight drop in information that reverses as correlations become stronger, increasing the advantage of the combinatorial code over the spike count code at high levels of correlation. Across correlation levels, the extra information from a code based on words versus counts is greater than or approximately equal to that found in the independent population.  Thus, our conclusions about the opportunities for combinatorial coding are robust across a wide range of correlation strengths, including those observed experimentally, which are usually in the range of 0.1 to 0.2$^\textrm{25,28-33}$. We conclude that combinatorial codes neither require exotic correlations among neurons, nor are they disrupted by the modest levels of correlation consistent with available data. 

\begin{figure}
\begin{center}
\leavevmode
\hbox{
{\includegraphics*[width=3in]{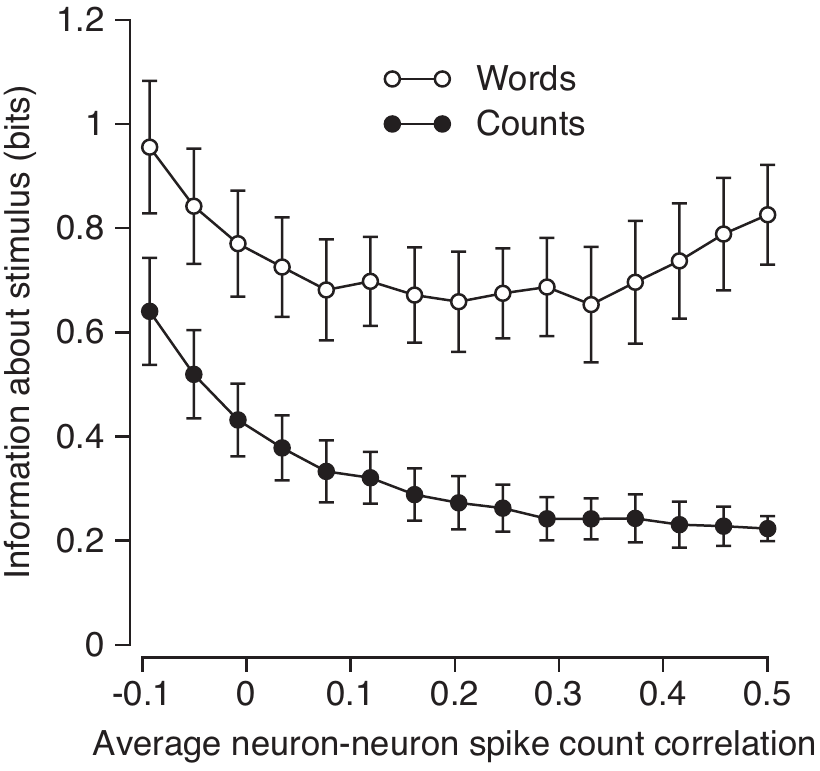}}
}
\caption{Impact of neuron-neuron correlations on coding based on population patterns of spikes and silence verus spike counts. The values on the x-axis are the calculated correlations between pairs of separately sampled units after setting up correlated populations using Equations 11-13. Filled and open circles show information about the stimulus from counts versus patterns of spiking and silence, respectively. Data are shown as means and standard deviations across 50 groups of $N=10$ cells drawn from our experimental sample of 36 neurons. }
\end{center}
\end{figure}

\noindent 

\noindent \textbf{\textit{Discussion}}

 For many years, the discussion of neural coding could be summarized along an axis that had two endpoints: ``rate'' vs. ``timing'' codes$^\textrm{34-36}$. There is no question that neurons respond to sensory inputs by changing the rate at which they generate action potentials, and many therefore believe that rate coding is firmly established.  In contrast, codes based on the timing of spikes, whether in sequence from a single neuron or across a population, have been viewed as more speculative. Our goal in this paper has been to replace the ``rate vs. timing'' debate with a sharper set of questions about the nature of the neural code, and to suggest that these questions have surprising answers.

 Our argument begins with experimental support for a purely conceptual point:  ``rate,'' as it is defined experimentally, cannot be a symbol in a code.  Symbols or code words must be observable directly in single trials, as with the letters or words in text. In contrast, rate refers to the \textit{probability} of generating spikes and hence is by definition an average property of spike trains over many trials.  The point that rate cannot be a symbol in the code would be pedantic if neurons operated in a regime where they generate large numbers of spikes before the rate has a chance to change significantly.  In this limit, rate can be estimated by counting spikes so that a ``spike count'' code becomes a reasonable surrogate for a ``rate code''.  But real neurons do not operate in this limit.  For example, cells in MT provide most of their information about motion direction with just a few spikes$^\textrm{11}$, and estimates of the underlying rate based on counting spikes in reasonable time windows are both uninformative (Figure 1d) and systematically in error (Figure 1c).  These problems are not ameliorated by averaging across a population of neurons with similar direction selectivity (Figure 1c).

 The second step in our argument is to realize that even precise statements about the statistics of spike trains do not yield unique conclusions about the symbolic structure of the code.  Thus, even for neurons whose spike train statistics are completely described by their underlying time-varying rate, estimates of the underlying rate from single spike trains depend on the timing of individual spikes.  The intuition that Poisson statistics imply a code based on spike counts is valid only in the limit where rates are nearly time independent.  Again, this analytical result is strongly reflected in the operation of real neurons:  for cells in MT, in the exact timing of spikes within a reasonable window is related to fluctuations of 25\% in our best estimate of the underlying rate (Figures 1a and b).  The fact that spike timing is critical for rate estimation in practice, as well as in principle, requires revision of the usual formulation of  ``rate vs. timing'' arguments.  

 The third step in our argument arises from realizing that the more important issue in neural coding and decoding is not the patterns of spike timing in single neurons, but rather its parallel in the patterns of spikes and silence across a population of neurons. We found that patterns of spiking and silence across a neural population contain twice as much information about a sensory stimulus as does the spike count. The extra information arises from the diversity of dynamical response properties across a population of neurons that otherwise have very similar tuning curves (e.g., for the direction of motion in MT). Indeed, it is this diversity of response dynamics that limits the effectiveness of simple averaging strategies, and, as such, cortical populations are only \textit{nominally} redundant.  Because we are considering the information carried by patterns in a single small time window, these results are unaffected by correlations across time (Poisson vs. non-Poisson statistics), and we have checked that they are also robust against reasonable levels of correlation among pairs of cells. We refer to the code defined by population words as \textit{combinatorial}, because it depends in a critical way on the combinations of spikes and silence across the neural population. Indeed, for particular code words, the combined response of the population carries more information than would be expected by adding up the information carried by the responses of the individual cells (Fig. 6). 

 We have presented our arguments in the concrete context provided by the coding of visual motion in area MT of the primate cortex, but the results should be much more general.  Certainly the theoretical difficulties with the traditional formulation of the coding problem as rate vs. timing are completely general, as seen from Equation \ref{GrindEQ__3_}. The quantitative results on the magnitude of the extra information carried in a combinatorial code depend on the details of the neural population we are considering, but we emphasize that there is nothing extreme about the population of cells we have analyzed in detail.  While the presence of extra information in the combinatorial code does not mean that brain uses this information to guide behavior, it is crucial that what might have seemed like an exotic coding scheme does not in fact depend upon the existence of unusual structures in the spike trains, either of single neurons or of populations.  Rather, the possibility of combinatorial coding is a previously unappreciated consequence of well-known dynamic response properties of neural responses throughout the cortex. While these cells encode stimuli by changing their firing rates, the elementary symbols of this code are the combinations of individual spikes and silences across the population of cells.

\noindent 

\noindent \textbf{\textit{Methods}}

\noindent \textit{{Experimental methods}}

Experimental data have been published previously$^\textrm{11}$. To acquire these data, extracellular single-unit microeletrode recordings were made in 3 sufentanil-anesthestized, paralyzed monkeys (\textit{Macaca fasicularis}) according to a protocol that had been approved in advance by the \textit{Institutional Animal Care and Use Committee} at UCSF. Using random dot texture stimuli presented on a high-resolution analog oscilloscope display, we mapped receptive field location, determined the preferred direction and speed of the neuron under study, and sized the stimuli to maximally excite each neuron. The random dot texture was moved behind a stationary aperture, creating a moving stimulus at a fixed retinal location. 

Visual stimuli were presented in discrete trials. Each stimulus appeared and remained stationary for 256 ms, then stepped to a constant velocity for 256 ms, and was again stationary for 256 ms. A brief pause separated successive trials, and directions of motion were pseudorandomly interleaved. A typical experiment included 13 motion directions that spanned$\pm$90 degrees around the neuron's preferred direction in 15 degree increments. Each stimulus was presented up to 222 times. Spike times were recorded with 10 microsecond resolution. 

\noindent \textit{{Constructing a model population}}

From the independently recorded single unit responses, we constructed a model of the population response to a motion stimulus. To create a population with nominally redundant feature selectivity, we aligned all cells by their preferred direction. Then, we resampled the rasters of individual cells at $\Delta \tau =$ 8 ms resolution, labelling the occupancy of each time bin with a ``1'' if there had been one or more spikes in the time interval or a ``0'' if there had not. At this resolution, multiple spikes in a single bin were infrequent, occurring in fewer than 10\% of the spiking events we recorded. We then created binary population ``words'', defined as patterns of 1's and 0's, at each time point during the response by randomly drawing the  $N$ letters of each word from the collection of stimulus repetitions from all $N$ cells in our sample in the appropriate bin. Each neuron in our sample corresponded to a fixed position in the word, and we could construct many different words by random draws from the many repetitions of each stimulus for each neuron.  The probability of observing a particular word then was measured by estimating the frequency of occurrence of that pattern of  1's and  0's within the entire dataset. 

\noindent \textit{{Estimating information}}

To estimate the information carried by population words about the stimulus, we first computed the probability of observing particular $N$-neuron words from our dataset $P(\textbf{n}\equiv \{ n_{{\rm i}} \} )$, where $i$ labels the neurons, over all time and for all motion directions. The total entropy of the words is given by: 
\begin{equation} \label{GrindEQ__4_} 
S[P(\textbf{n})]=-\sum _\textbf{n} P(\textbf{n})\log_{2} P(\textbf{n}) 
\end{equation} 
where  is a label that indexes word identity. The probability of observing a word for a particular stimulus, $P(\textbf{n}|\theta,t)$, was estimated in a similar manner to $P(\textbf{n})$ but at particular time $t$ relative to the onset of motion in direction $\theta $. The entropy of the conditional distributions is given by: 
\begin{equation} \label{GrindEQ__5_} 
S[P(\textbf{n}|\theta,t)]=-\sum _\textbf{n} P(\textbf{n}|\theta,t)\log_{2} P(\textbf{n}|\theta,t). 
\end{equation} 
The average amount of information that words carry about the stimulus is given by the difference between the total entropy and the average noise entropy:
\begin{equation} \label{GrindEQ__6_} 
I_\textrm{words} =S[P(\textbf{n})]-\left(\frac{1}{13\cdot T} \sum _{t}\sum _{\theta }S[P(\textbf{n}|\theta ,t)]  \right) 
\end{equation} 
where $T$ represents the total number of time bins in the response and $\theta $ indexes the 13 motion directions. In a similar way, we can compute the information from counts using the same $P(\textbf{n})$ as before, but collapsing over words with the same number of spikes, such that $P(\textrm{count}=n)=\sum P(\textbf{n})$, where the sum runs over all words, $\textbf{n}$, with count equal to $n\equiv \sum _{{\rm i}=1}^{N} n_{{\rm i}} $. Similarly, $P(\textrm{count}=n|\theta,t)=\sum P(\textbf{n}|\theta,t)$.  With this $P(\textrm{count})$ in hand, we compute $I_\textrm{counts}$ in a completely analogous fashion to the calculation of $I_\textrm{words}$. 

In those cases where we generate samples of population words directly from observed spike trains, all entropy estimates were corrected for finite sampling effects by taking multiple random samples of fractions of the dataset and then performing a linear extrapolation to infinite sample size$^\textrm{37}$.  Errors in estimates were estimated by extrapolating the standard deviation of values computed from half the sample in the same manner. Because we ask \textit{only} about words formed from responses in a single time bin, correlations between time bins (and hence the question of whether the neurons are exactly Poisson) are irrelevant; as a test of our computations we created shuffled spike trains with exact Poisson statistics, and reproduced all of our results.

Calculations based on a model population sampled from the real data have a strong intuitive connection to experiment, motivating us to use the approach outlined above to make estimates of information based on real data.  However, we found that working with real data was unsatisfactory in some ways; information rates converge only for a very large number of samples, which becomes increasingly cumbersome as the number of neurons, $N$, exceeds 16. Because the statistics we wish to reproduce in this model population are just the time dependent, experimentally observed firing rates for each neuron, it was possible to do the same analysis after simply calculating $P(\textrm{word})$ using Equation 10 (see below).  This approach works because the spike rate of each model neuron, at each moment of time, is determined by our experimental data with small error bars, and hence there are no free parameters in the construction of our model population. We checked that this approach yielded the same answers as the data-based approach for values of $N$ where the calculations were tractable. In those cases where we computed word probabilities directly from Equation \ref{GrindEQ__10_}, we propagated the errors in measured firing rates to obtain errors in the derived information measures.

To create the model populations with identical time varying firing rates or directional tuning curves used to generate the data in Figure 3, we used each neuron in turn as a template.  For each group of 10 model cells, we randomly chose another neuron from the population, using its tuning curve, $\bar{r}(\theta)^*$ (the bar indicates a time average), and the shape of its temporal modulations in rate at the preferred orientation, $r(t)^*=r(t,\theta=0)/\bar{r}(\theta=0)$, to serve as a template for fixing the tuning or firing rate dynamics of the group.  To fix the tuning of the population, we allowed each cell to retain its own $r(t)$, but rescaled each curve by a constant factor which forced the cell's tuning curve to follow $\bar{r}(\theta)^*$, such that $r(t,\theta)=[r(t,\theta)/\bar{r}(\theta)]\cdot \bar{r}(\theta)^*$.  To fix the firing rate dynamics, each cell retained its own directional tuning curve, but temporal dynamics were set by the template, $r(t)^*$, so that $r(t,\theta)=r(t)^*\bar{r}(\theta)$.

\noindent \textit{{Spike-silence synergy}}

To measure the synergy between spikes and silences in our population words, we simply took the difference between the stimulus information that word captured and the sum of the information from each component spike and silence$^\textrm{20,38}$, 
\begin{equation} \label{GrindEQ__7_} 
I_\textrm{synergy} =I(\{ n_{i} \} )-\sum _{i}I(n_{i} ).  
\end{equation} 
The stimulus information, $I$, is computed as in Brenner et al., 2000$^\textrm{38}$, and is given by: 
\begin{equation} \label{GrindEQ__8_} 
I(\textrm{stimulus})=\frac{1}{T} \int _{0}^{T}dt\frac{r(t)}{\left\langle r(t)\right\rangle _{t} }  \log _{2} \left(\frac{r(t)}{\left\langle r(t)\right\rangle } _{t} \right) 
\end{equation} 
where $r(t)$ is the modulation of the rate for a given event, the occurrence of a given work or a spike or silence from a particular cell.

\noindent \textit{{Constructing a correlated population}}

We continue to work in small time windows, of duration $\Delta \tau $, such that the response of each neuron $i$ consists either of a spike $(n=1)$ or silence $(n=0)$. Then if all cells respond independently, we can write the probability distribution for the population's response $\{ n_i\}$ at some moment of time $t$ in the form: 
\begin{equation} \label{GrindEQ__9_} 
P(\{ n_i \} |t) = \prod_{i=1}^N\frac{[q_i(t)]^{n_i}}{1+q_i(t)}
\end{equation} 
where $q_i(t)$ denotes the probability of a spike from the $i^\textrm{th}$ neuron at time $t$, and in the limit $\Delta\tau \rightarrow 0$ we can identify the time dependent firing rate of each neuron $r_i(t) = q_i(t)/(\Delta\tau)$. Equation \ref{GrindEQ__9_} can be rewritten as: 
\begin{equation} \label{GrindEQ__10_} 
P( \{ n_i \} |t) =\frac{1}{Z(t)}\exp \left [ \sum_{i=1}^N \phi_i(t) n_i  \right ]
\end{equation} 
where $Z(t)$ is a normalization constant and $\phi _{{\rm i}} (t)=\ln q_{{\rm i}} (t)$. This form suggests that we can add correlations among neurons by adding an explicit term to the exponential that couples the responses of the different cells: 
\begin{multline} \label{GrindEQ__11_} 
P( \{ n_i \} |t) =\\ \frac{1}{Z(t)}\exp \left [ \sum_{i=1}^N \phi_i(t) n_i +\frac{J}{2}\sum_{i=1}^N\sum_{j\neq i}n_i n_j \right ] 
\end{multline} 
In the independent model, there are no correlations between the responses $n_{{\rm i}} $ and $n_{{\rm j}} $ once we know the stimulus, while Equation \ref{GrindEQ__11_} predicts that there will be non-zero correlations; for small $J$, the strength of these correlations is proportional to $J$. In fact, Equation \ref{GrindEQ__11_} is the least structured, or maximum entropy model that generates some average level of correlations among all the pairs of cells$^\textrm{39-41}$.

To produce a model population of neurons with an average pairwise coupling, $J$, which respects each cell's average firing rate as a function of time $r_{i} (t)$, we need to solve for the $\phi _{{\rm i}} (t)$'s in Equation \ref{GrindEQ__11_}, subject to the constraints: 
\begin{equation} \label{GrindEQ__12_} 
\sum_\textbf{n} n_k(t)P(\textbf{n}|t)=r_k\Delta\tau 
\end{equation} 
where the $r_k(t)$'s are measured single cell firing rates, averaged over a small time window, $\Delta \tau =$ 8 ms. Since the cells are not coupled in time, we can solve for the fields at each time point independently. We have an analytical solution for the fields with $J=0$ and we can proceed from this solution using perturbation theory, from which we obtain an equation relating small changes in  to their effect on the fields, $\phi $:
\begin{equation} \label{GrindEQ__13_} 
\Delta\phi_i = \left( \frac{\Delta J}{2}\sum_{\alpha\neq\beta} \langle n_\alpha n_\beta \rangle \langle n_k \rangle - \langle n_\alpha n_\beta n_k\rangle \right ) \chi_{ik}^{-1} 
\end{equation} 
where $\alpha $ and $\beta $ index neurons in the group  cells and $\chi $ is the connected part of the two-point correlation function, $\chi_{ik} =\langle n_{{\rm i}} n_{{\rm k}} \rangle -\langle n_{{\rm i}} \rangle \langle n_{{\rm k}} \rangle $. We solve for the fields at very small increments, $\Delta J/J=0.001$, checking satisfaction of the constraints on the firing rates at each step. This perturbative approach is fast, but accumulates errors. To correct for the accumulated errors we perform local function minimization whenever the fractional error in the single cell rates exceeds $10^{-8}$, and then return to the perturbative stepping until the error bound is again reached.

Once we create a model population response, we sum the spike counts across the full time window of the response to motion, and compute the correlation coefficients between counts in all pairs of cells in our model population The mean of these coefficients provides an index for the overall strength of the correlations. Experimentally, for neurons in MT, the correlation coefficients are in the range from 0.1 to 0.2$^\textrm{25,28-33}$, which corresponds to $J=0.11$ to 0.16 in our models.

\noindent \textbf{Acknowledgments:} Supported by NIH Grant EY017210, the Howard Hughes Medical Institute, the Life Sciences Research Foundation, and the Swartz Foundation.

\noindent \textbf{REFERENCES}\textit{ }

\noindent [1] Adrian, E. D. The impulses produced by sensory nerve endings: Part I. \textit{ J. Physiol. (Lond)} \textbf{ 61,} 49-72 (1926). 

\noindent [2] Adrian, E. D. \& Zotterman, Y. The impulses produced by sensory nerve endings: Part II. The response of a single end organ. \textit{ J. Physiol. (Lond)} \textbf{ 61,} 151-171 (1926). 

\noindent [3] Adrian, E. D. \& Zotterman, Y. The impulses produced by sensory nerve endings: Part III. Impulses set up by touch and pressure. \textit{ J. Physiol. (Lond)} \textbf{ 61,} 465-483 (1926). 

\noindent [4] Hartline, H. K. The receptive fields of optic nerve fibers. \textit{Am. J. Physiol.} \textbf{130}, 690-699 (1940). 

\noindent [5] Barlow, H. B. Summation and inhibition in the frog's retina. \textit{ J. Physiol. (Lond)} \textbf{ 119,} 69-88 (1953). 

\noindent [6] Kuffler, S. W. Discharge patterns and functional organization of mammalian retina. \textit{ J. Neurophysiol.} \textbf{16,} 37-68 (1953). 

\noindent [7] Hubel, D. H. \& Wiesel, T. N. Receptive fields, binocular interaction and functional architecture in the cat's visual cortex. \textit{ J. Physiol. (Lond.)} \textbf{160,} 106-154 (1962). 

\noindent [8] Mountcastle, V. B. Modality and topographic properties of single neurons of cat's somatic sensory cortex. \textit{ J. Neurophysiol.} \textbf{20,} 408-434 (1957).

\noindent [9] Mountcastle, V. B. The columnar organization of the neocortex. \textit{ Brain} \textbf{ 120,} 701-722 (1997). 

\noindent [10] Rieke, F., Warland, D., de Ruyter van Steveninck, R. R. \& Bialek, W.\textit{ Spikes: Exploring the Neural Code.} (MIT Press, Cambridge, MA) (1997). [10] [Perkel and Bullock 1968] 

\noindent [11] Osborne, L. C., Bialek, W. \& Lisberger, S. G. Time course of information about motion direction in visual area MT of macaque monkeys. \textit{ J. Neurosci.} \textbf{24,} 3210-3222 (2004). 

\noindent [12] Perkel, D. H. \& Bullock, T. H. Neural coding. \textit{ Neurosci. Res. Prog. Sum.} \textbf{3,} 405-527 (1968). 

\noindent [13] Reich, D. S., Mechler, F. \& Victor, J. D. Independent and redundant information in nearby cortical neurons. \textit{ Science} \textbf{ 294,} 2566-2568 (2001). 

\noindent [14] Newsome, W. T., Wurtz, R. H., Dursteler M. R. \& Mikami, A. Deficits in visual motion processing following ibotenic acid lesions of the middle temporal visual area of the macaque monkey. \textit{ J. Neurosci. }\textbf{5,} 825-840 (1985).

\noindent [15] Osborne, L. C.  \& Lisberger, S. G. Spatial and temporal integration of stochastic motion direction signals in primate pursuit eye movements. Program No. 715.24. Abstract Viewer and Itinerary Planner. Washington, DC: Society for Neuroscience, (2007) Online

\noindent [16] Osborne, L. C., Hohl, S. S., Bialek, W. \& Lisberger, S. G., Time course of precision in smooth pursuit eye movements of monkeys. \textit{ J. Neurosci.} \textbf{ 27,} 2987-2998 (2007). 

\noindent [17] Shamir, M. \& Sompolinsky, H. Implications of neuronal diversity on population coding.  \textit{Neural Comput.} \textbf{18}, 1951-86 (2006).

\noindent [18] de Ruyter van Steveninck, R. R. \&  Bialek, W. Real-time performance of a movement sensitive neuron in the blowfly visual system: Coding and information transfer in short spike sequences. \textit{ Proc. R. Soc. London Ser. B} \textbf{234}, 379-414 (1988). 

\noindent [19] Osborne, L. C., Lisberger, S. G. \& Bialek, W. A sensory source for motor variation. \textit{ Nature} \textbf{ 437,} 412-416 (2005). 

\noindent [20] Schneidman, E., Puchalla, J. L., Harris, R. A., Bialek W. \& Berry II, M. J. Synergy from silence in a combinatorial neural code. q-bio.NC/0607017 (2006). 

\noindent [21] Johnson, K. O. Sensory discrimination: decision process. \textit{J. Neurophysiol.} \textbf{43}, 1771-1792 (1980).

\noindent [22] Britten, K. H., Shadlen, M. N., Newsome, W. T. \& Movshon, J. A.  The analysis of visual motion: a comparison of neuronal and psychophysical performance. \textit{J. Neurosci.} \textbf{12},4745-4765 (1992).

\noindent [23] Seung, H.S. \& Sompolinsky, H. Simple models for reading neuronal population codes.  \textit{Proc. Natl. Acad. Sci.} USA \textbf{90}, 10749-10753 (1993).

\noindent [24] Abbott, L. F. Decoding neuronal firing and modeling neural networks.  \textit{Q. Rev. Biophys.} \textbf{27}, 291-331 (1994).

\noindent [25] Zohary, E., Shadlen, M. N. \& Newsome, W. T. Correlated neuronal discharge rate and its implications for psychophysical performance. \textit{Nature} \textbf{370}, 140-143 (1994).

\noindent [26] Shadlen, M. N., Britten, K. H., Newsome, W. T. \& Movshon, J. A. A computational analysis of the relationship between neuronal and behavioral responses to visual motion.  \textit{J. Neurosci.} \textbf{16}, 1486-1510 (1996).

\noindent [27] Abbott, L. F. \& Dayan, P. The effect of correlated variability on the accuracy of a population code.  \textit{Neural Comput.} \textbf{11}, 91-101 (1999).

\noindent [28] Van Kan, P. L. E., Scobey, R. P. \& Gabor A. J. Response covariance in cat visual cortex.  \textit{Exp. Brain Res.} \textbf{60}, 559-563 (1985).

\noindent [29] Gawne ,T. J. \& Richmond, B. J. How independent are the messages carried by adjacent inferior temporal cortical neurons?  \textit{J. Neurosci.} \textbf{13}, 2758-2771 (1993).

\noindent [30] Lee, D., Port, N. L., Kruse, W. \& Georgopoulos, A.P. Variability and correlated noise in the discharge of neurons in motor parietal areas of the primate cortex. \textit{J. Neurosci.} \textbf{18},1161-1170 (1998).

\noindent [31] Bair, W., Zohary, E. \& Newsome, W. T. Correlated firing in macaque visual area MT: time scales and relationship to behavior\textit{.  J. Neurosci.} \textbf{21}, 1676-1697 (2001).

\noindent [32] Averbeck, B. B. \& Lee, D. Neural noise and movement-related codes in the macaque supplementary area.  \textit{J. Neurosci.} \textbf{23},7630-7641 (2003).

\noindent [33] Kohn, A. \& Smith, M.A. Stimulus dependence of neuronal correlation in primary visual cortex of the macaque\textit{.  J. Neurosci. }\textbf{25}, 3661-3673 (2005).

\noindent [34] Theunissen, F. \& Miller, J. P.  Temporal encoding in nervous systems: a rigorous definition. \textit{J. Comput. Neurosci.} \textbf{2}, 149-162 (1995).

\noindent [35] Van Rullen, R. \& Thorpe, S. J. Rate coding versus temporal order coding: what the retinal ganglion cells tell the visual cortex. \textit{Neural Comput.} \textbf{13}, 1255-1283 (2001).

\noindent [36] Dayan, P. \& Abbott, L. F. Theoretical Neuroscience: Computational and Mathematical Modeling of Neural Systems. (The MIT Press, Cambridge, MA, USA) (2001).

\noindent [37] Strong, S. P., Koberle, R., de Ruyter van Steveninck, R. R. \& Bialek, W. Entropy and information in neural spike trains. \textit{ Phys. Rev. Lett.} \textbf{ 80,} 197-200 (1998).

\noindent [38] Brenner, N., Strong, S. P., Koberle, R., Bialek, W. \& de Ruyter van Steveninck, R. R. Synergy in a neural code. \textit{ Neural Comput.} \textbf{ 12,} 1531-1552 (2000). 

\noindent [39] Jaynes, E. T. Information theory and statistical mechanics. \textit{ Phys. Rev.} \textbf{106,} 62-79 (1957).

\noindent [40] Schneidman, E., Still, S., Berry II, M. J. \& Bialek, W. Network information and connected correlations. \textit{ Phys. Rev. Lett.} \textbf{ 91,}  238701-231704 (2003); physics/0307072. 

\noindent [41] Schneidman, E., Berry II, M. J., Segev, R. \& Bialek, W. Weak pairwise correlations imply strongly correlated network states in a neural population. \textit{ Nature} \textbf{ 440,} 1007-1012 (2006); q-bio.NC/0512013. 

\end{document}